\documentclass[a4paper]{article}

\usepackage{url}
\PassOptionsToPackage{hyphens}{url}

\usepackage{geometry}
\usepackage{graphicx}   
\usepackage{setspace}
\usepackage{amsmath, amssymb}
\usepackage{titlesec}
\usepackage{fancyhdr}
\usepackage{multicol}
\usepackage{parskip}
\usepackage{indentfirst}
\usepackage{etoolbox}
\usepackage{caption}
\usepackage{cite}
\usepackage{xcolor}

\usepackage{xurl} 
\usepackage[english]{babel}  % Hoặc
\usepackage{algorithm}
\usepackage{algpseudocode}

\usepackage{subcaption}
\usepackage{float}
\usepackage{placeins}

\usepackage[breaklinks=true]{hyperref}
\usepackage[T1]{fontenc}
\usepackage[utf8]{inputenc}

\titleformat{\section}{\centering\large\scshape}{\thesection}{1em}{}
\titleformat{\subsection}{\normalfont\large\itshape}{\thesubsection}{1em}{}
\titleformat{\subsubsection}{\normalfont\itshape}{\thesubsubsection}{1em}{}
\titlespacing{\section}{0pt}{6pt}{6pt}
\titlespacing{\subsection}{0pt}{6pt}{6pt}
\titlespacing{\subsubsection}{0pt}{6pt}{6pt}

\title{
    \textbf{A Cost-Optimization Model for EV Charging Stations Utilizing Solar Energy and Variable Pricing}
    
}
\author{} % Ẩn danh tác giả nếu cần
\date{}   % Không hiển thị ngày

\renewcommand{\thesection}{\Roman{section}}
\renewcommand{\thesubsection}{\thesection.\Alph{subsection}}
\begin{document}
\maketitle
\vspace{-1.5cm}

\begin{multicols}{3}
    \centering
    \textbf{An Nguyen}\\
    \textit{College of Engineering and Computer Science, VinUniversity}\\
	\vfill

    \columnbreak

    \textbf{Hung Pham}\\
    \textit{College of Engineering and Computer Science, VinUniversity}\\
    \vfill

    \columnbreak

    \textbf{Cuong Do}\\
    \textit{College of Engineering and Computer Science, VinUniversity}\\
    \textit{Center for Environmental Intelligence (CEI), VinUniversity}\\
    \vfill
\end{multicols}

% ABSTRACT IN SINGLE COLUMN
\section*{Abstract}

This paper presents a cost optimization framework for electric vehicle (EV) charging stations that leverages on-site photovoltaic (PV) generation and explicitly accounts for electricity price uncertainty through a Bertsimas--Sim robust formulation. The model is formulated as a linear program that satisfies vehicle energy demands, respects charging and grid capacity constraints, and minimizes procurement cost.  Evaluations on real charging data from the Caltech ACN dataset show average savings of about 12\% compared to a first-come--first-served baseline, with peak monthly reductions up to 19.2\%. A lightweight sensitivity analysis indicates that a modest $\sim$5\% increase in nominal cost can reduce worst-case exposure by 14\%. Computational tests confirm real-time feasibility, with instances of up to 50 concurrent EVs solved in under 5 seconds on a standard laptop. The proposed method provides a practical, grid-friendly, and scalable solution for future EV charging operations.

\begin{multicols}{2} % BEGIN TWO-COLUMN LAYOUT

\setlength{\columnsep}{0.5cm}
\setlength{\parindent}{0.5cm}
\setlength{\parskip}{6pt}
\singlespacing

\section{Introduction}
Climate change is one of the greatest challenges of our time, and Vietnam’s commitment to achieving net-zero emissions by 2050 \cite{1} underscores the nation’s determination to play a leadership role in global sustainability. However, realizing this vision requires tackling significant hurdles—particularly in the transport sector, which currently accounts for about 10.8\% of total emissions and is projected to increase its CO$_2$ output by 6–7\% annually \cite{2}. Accelerating the adoption of electric vehicles (EVs) has emerged as a pivotal strategy to curb these emissions and mitigate the impact of a rapidly expanding transport sector \cite{3}.

A key indicator of Vietnam’s EV momentum is the rise of VinFast, the country’s first domestic EV manufacturer. Launched in late 2021, VinFast quickly established its inaugural line of electric cars and, by 2024, had sold over 51,000 vehicles \cite{4,5}. While this rapid growth reflects strong consumer demand and technological promise, it also exposes critical infrastructure gaps. Insufficient charging stations \cite{6} and an aging urban grid system, already under strain, threaten to stall progress and reduce the overall effectiveness of EVs in cutting emissions. To address these obstacles and sustain EV growth, Vietnam must make substantial investments in renewable energy, especially solar, which can help meet surging electricity demands while advancing the nation’s net-zero ambition \cite{7}.

Smart Charging is a compelling research topic with numerous subproblems that many researchers are actively addressing. These subproblems include Demand Forecasting, Charging Schedule Optimization, Power Allocation, Charging Infrastructure and Grid Connection, Communication and Payment, and Security and Privacy. Various advanced methodologies have been applied to tackle these challenges.

In \cite{8}, a hybrid model combining Deep Reinforcement Learning (DRL) with the Quantum-Inspired Genetic Algorithm (QIGA) was proposed to address the Optimal Power Flow (OPF) problem in hybrid renewable energy systems. In \cite{9}, a quadratic programming and convex problem formulation approach was introduced for smart charging in a single household equipped with an electric vehicle and an energy storage system. Linear Programming and Robust Optimization were employed in \cite{10} to handle uncertainties in electricity prices and energy demand. In \cite{11}, the mean clustering algorithm and queuing theory were utilized to determine the optimal location and capacity of charging stations.

Our approach is based on linear programming, a mathematical optimization method used to minimize costs by defining a linear objective function and a set of linear constraints that represent the system’s operational limits. To improve the model’s resilience in real-world scenarios, we incorporate a robust optimization framework that accounts for uncertainty in electricity prices using the Bertsimas–Sim methodology. The remainder of this paper is structured as follows: Section 2 introduces the key symbols and parameters. Section 3 then presents the base optimization model formulation. Building upon this, Section 4 details the complete proposed methodology, which includes the robust optimization framework and an online algorithm for real-time implementation. Section 5 presents a comprehensive evaluation of our framework, including performance comparisons against a baseline, a sensitivity analysis, and computational tests. Finally, Section 6 concludes the paper with a summary of findings, while Section 7 discusses the study's limitations and outlines directions for future work.

\section{Symbols and Parameters}

\begin{itemize}
\item $T$: Number of time steps in a day (e.g., $T = 24$ hours).
\item $\Delta t$: Length of each time step (hours).
\item $N$: Number of electric vehicles (EVs).
\item $p_t^{\text{grid}}$: Electricity price from the grid at time $t$ (Euro/kWh).
\item $R_t$: Solar energy output at time $t$ (kW).
\item $\bar{R}_t$: Maximum achievable solar energy output at time $t$ (kW).
\item $s_i$: Maximum charging power provided by the charging station for EV $i$ (kW).
\item $A_{t,i}$: Matrix representing the charging time of vehicle $i$ by hour.
\item $C_{\text{grid}}$: Grid capacity limit (kW).
\item $\eta$: Charging efficiency of the EV (in the range $(0,1]$).
\item $L_i$: Minimum required energy (kWh) that EV $i$ needs when leaving the station.
\item $\mathcal{T}_i$: Set of time points when EV $i$ is present at the station (based on $A_{t,i}$).
\end{itemize}

\section{Optimization Model}
Several straightforward solutions have been considered for managing EV charging, such as the “First Come, First Served” (FCFS) principle, which guarantees equity by servicing vehicles in the sequence of their arrival. However, FCFS lacks consideration of critical factors such as electricity pricing, real-time energy demand, and overall impact on the power grid. This limitation can lead to inefficient energy allocation, increased operational costs, and potential grid instability, especially during peak demand periods.

\noindent
\begin{minipage}{\columnwidth}
\centering
\includegraphics[width=1\textwidth]{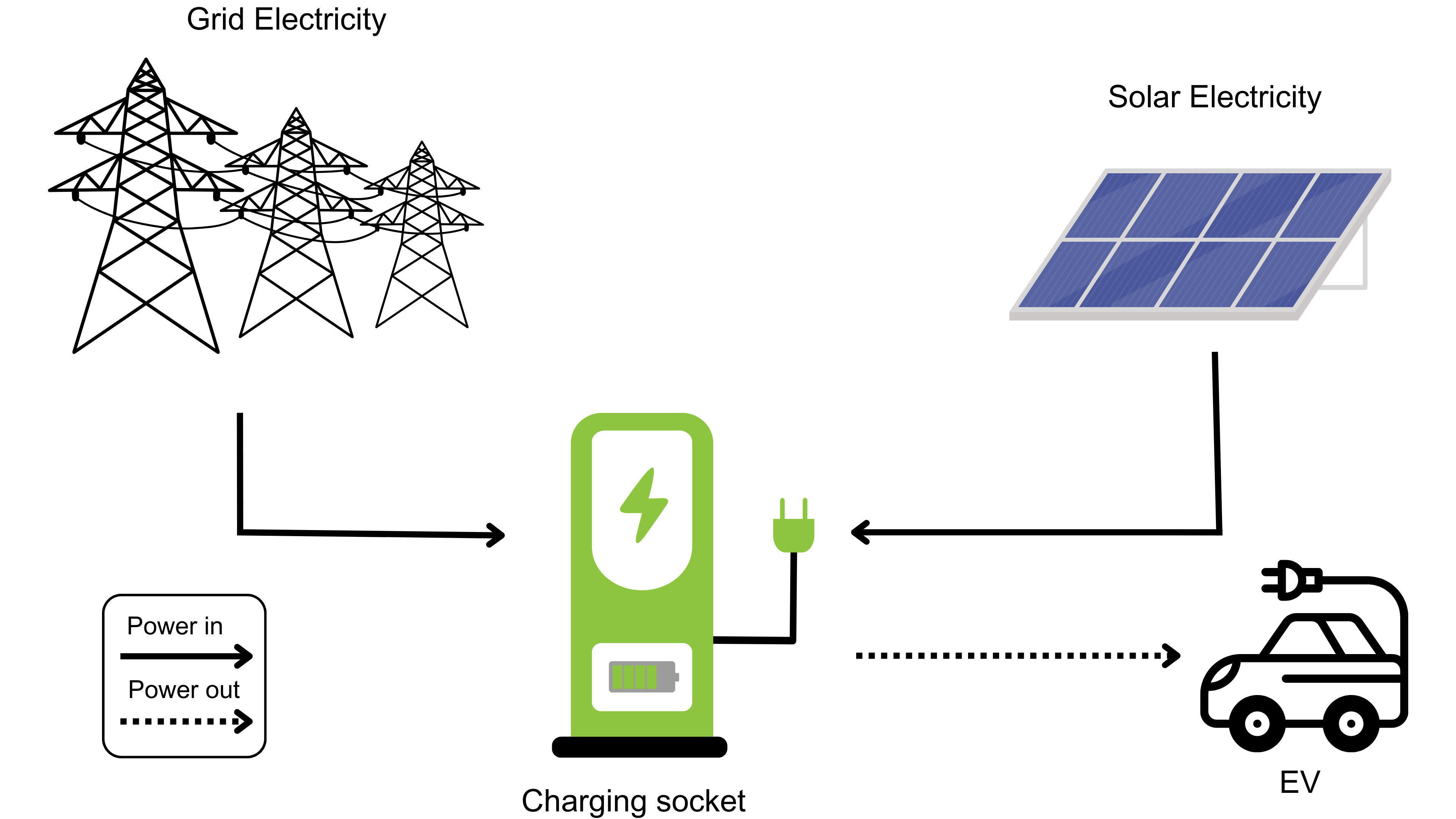}
\captionof{figure}{On-Site Solar-Integrated Charging Station Model}
\end{minipage}

Figure 1 illustrates the power flow diagram of charging stations with solar-integrated. The charging socket receives energy from both the solar panel and the grid, and then delivers it to the EVs. In this model, we employ a Linear Programming approach to minimize net electricity consumption costs, which occur only when the charging demand exceeds the available solar renewable energy. 

\noindent\textbf{Decision Variables:} 
\begin{itemize}
    \item $Y_{i,t} \ge 0$: Charging power (kW) of EV $i$ at time $t$.
    \item $S_t^+ \ge 0$: Auxiliary variable representing the positive part of the net load at time $t$, that is:
    \[
    S_t^+ = \max\Bigl\{\sum_{i=1}^{N} Y_{i,t} - R_t,\; 0\Bigr\}.
    \]
    \item $R_t$: The amount of solar energy used at time $t$.
\end{itemize}

\noindent\textbf{Objective Function:} 
\[
\min_{Y,\,S^+,\,R} \quad \sum_{t=1}^{T} p_t^{\text{grid}} \, S_t^+ \, \Delta t,
\]
where \( S_t^+ \) is used to linearize the expression \(\bigl(\sum_{i=1}^{N} Y_{i,t} - R_t\bigr)\).

\noindent The solar contribution \( R_t \) is computed as follows:
\[
R_t = A_{\mathrm{pv}} \, \frac{G(t)}{1000} \, \eta_{\mathrm{pv}},
\]
where:
\begin{itemize}
    \item \(A_{\mathrm{pv}}\) indicates the area of the PV panels in m$^2$,
    \item \(G(t)\) indicates the solar irradiance in W/m$^2$,
    \item \(\eta_{\mathrm{pv}}\) is the efficiency of the PV panels.
\end{itemize}

\noindent\textbf{Constraints:}

1. Each EV $i$ must be charged with at least the minimum required energy $L_i$ throughout its available period $T_i$. Here, $\eta$ is the charging efficiency, $Y_{i,t}$ is the charging power at time $t$, and $\Delta t$ is the length of each time interval. This requirement can be expressed as:
\[
\eta \sum_{t \in T_i} Y_{i,t}\,\Delta t \ge L_i, \quad \forall i.
\]

2. The charging power of each EV at time $t$ must not exceed the maximum power limit $s_i$ that the station provides. However, the actual available power is scaled by the fraction of the time interval during which EV $i$ is connected, denoted by $A_{t,i}$. For example, if the EV is connected only for 10 minutes in an hour (i.e., $A_{t,i}=\dfrac{10}{60}$), then the maximum available charging power becomes $s_i \times \dfrac{10}{60}$. When the EV is not connected (i.e., $A_{t,i}=0$), no charging power is provided, ensuring that $Y_{i,t}=0$. This is modeled by:
\[
Y_{i,t} \le s_i \, A_{t,i}, \quad \forall i,\,\forall t.
\]

3. The total charging power from all EVs at time $t$ must not exceed the maximum grid capacity $C_{\text{grid}}$ after subtracting the renewable energy $R_t$:
\[
\sum_{i=1}^{N} Y_{i,t} - R_t \le C_{\text{grid}}, \quad \forall t.
\]

4. The variable $S_t^+$ represents the additional power to be purchased from the grid if the total charging power exceeds $R_t$. When renewable energy sufficiently supplies the EVs, $S_t^+$ may be zero. It is ensured that the power purchased from the grid cannot be negative:
\[
S_t^+ \ge 0, \quad \forall t.
\]

5. Finally, the solar energy $R_t$ cannot exceed its maximum available limit $\bar{R}_t$ at each time and it must larger or equal than 0
\[
0 \le R_t \le \bar{R}_t,\quad \forall\, t.
\]

\noindent\textbf{Database:}\\
\indent We obtained detailed charging session information from 2018 to 2019 from ACN Data---a public dataset on electric vehicle (EV) charging collected through a collaboration between the PowerFlex System and the California Institute of Technology (Caltech)~\cite{12}. This dataset comprises detailed information on EV charging sessions at two distinct locations: the Caltech campus and the Jet Propulsion Laboratory (JPL) campus. The JPL site is representative of workplace charging, whereas Caltech represents a hybrid of workplace and public charging.

We selected hourly electricity price data for Spain corresponding to the same period as the charging data, sourced from Ember---European Wholesale Electricity Price Data~\cite{13}.

Hourly irradiance data were obtained from the EU Science Hub~\cite{14} for the same period. We assume that $G_t$ represents $G(i)$ [W/m\textsuperscript{2}]---the global in-plane irradiance with a slope of $36^\circ$ and an azimuth of $0^\circ$---with Madrid, Spain, chosen as the representative location.

\noindent
\begin{minipage}{\columnwidth}
\centering
\captionof{table}{Constant values} 
\label{table:KeyInputData}
\begin{tabular}{|l|l|}
\hline
\textbf{Parameter} & \textbf{Value} \\
\hline
$C_{\text{grid}}$ & 300 kW \\
$\eta$            & 0.9 \\
$\eta_{\mathrm{pv}}$ & 0.2 \\
$A_{\mathrm{pv}}$ & 80 m$^2$ \\
$s_i$ (Caltech)   & 86 kW \\
$s_i$ (JPL)       & 37.5 kW \\
\hline
\end{tabular}
\end{minipage}

\section{Proposed Methodology }

To account for uncertainty in the electricity price, we model the grid price at time \(t\) as
\[
p_t^{\text{grid}} = \hat{p}_t + \Delta p_t,
\]
where \(\hat{p}_t\) is the nominal electricity price and \(\Delta p_t\) represents the deviation from this nominal value. We assume that the deviation is bounded as
\[
|\Delta p_t| \le \overline{\Delta p}_t, \quad \forall t.
\]
To avoid an overly conservative solution—i.e., assuming that every time period experiences the maximum deviation simultaneously—we adopt a budget of uncertainty \(\Gamma\). The uncertainty set for \(\Delta p = (\Delta p_1, \dots, \Delta p_T)\) is defined as
\[
\mathcal{U} = \left\{ \Delta p \in \mathbb{R}^T : |\Delta p_t| \le \overline{\Delta p}_t,\quad \sum_{t=1}^{T} \frac{|\Delta p_t|}{\overline{\Delta p}_t} \le \Gamma \right\}.
\]

The nominal electricity cost from the grid is given by
\[
\sum_{t=1}^{T} p_t^{\text{grid}} \, S_t^+\, \Delta t = \sum_{t=1}^{T} \hat{p}_t\, S_t^+\, \Delta t,
\]
where \(S_t^+\) represents the purchased electricity (in kW) from the grid at time \(t\). Under uncertainty, the robust counterpart of the objective becomes a min–max formulation:
\[
\min_{Y,\,S^+,\,R} \; \max_{\Delta p \in \mathcal{U}} \; \sum_{t=1}^{T} \bigl(\hat{p}_t + \Delta p_t\bigr)\, S_t^+\, \Delta t.
\]
This expression can be decomposed into the nominal cost and the additional cost resulting from the uncertainty:
\[
\sum_{t=1}^{T} \hat{p}_t\, S_t^+\, \Delta t \;+\; \max_{\Delta p \in \mathcal{U}} \; \sum_{t=1}^{T} \Delta p_t\, S_t^+\, \Delta t.
\]

\subsection{Bertsimas--Sim Reformulation}

We now reformulate the inner maximization problem:
\[
\max_{\Delta p \in \mathcal{U}} \; \sum_{t=1}^{T} \Delta p_t\, S_t^+\, \Delta t,
\]
subject to
\[
\begin{array}{rl}
|\Delta p_t| &\le \overline{\Delta p}_t, \quad \forall t, \\
\displaystyle \sum_{t=1}^{T} \frac{|\Delta p_t|}{\overline{\Delta p}_t} &\le \Gamma.
\end{array}
\]
Here, the terms \(S_t^+\) and \(\Delta t\) are treated as fixed parameters. Following the approach of Bertsimas and Sim in ~\cite{15}, we introduce an auxiliary scalar variable \(\lambda \ge 0\) and auxiliary variables \(\mu_t \ge 0\) for all \(t=1,\ldots,T\). The worst-case additional cost due to price deviation is then equivalently expressed as
\[
\Gamma\,\lambda + \sum_{t=1}^{T} \mu_t,
\]
subject to the dual feasibility constraints
\[
\mu_t \ge \overline{\Delta p}_t\, S_t^+\, \Delta t - \lambda, \quad \forall t.
\]

\subsection{Final Robust Optimization Model}

Incorporating the Bertsimas--Sim reformulation into the full model, the robust optimization problem is given by:
\[
\begin{array}{rlcl}
\displaystyle \min_{Y,\,S^+,\,R,\lambda,\mu} & \displaystyle \sum_{t=1}^{T} \hat{p}_t\, S_t^+\, \Delta t \;+\; \Gamma\,\lambda + \displaystyle \sum_{t=1}^{T} \mu_t & & \\[2mm]
\text{s.t.} & \mu_t \ge \overline{\Delta p}_t\, S_t^+\, \Delta t - \lambda, & \quad & \forall\, t, \\[1mm]
            & \lambda \ge 0,\quad \mu_t \ge 0, & \quad & \forall\, t, \\[2mm]
            & \eta \displaystyle \sum_{t \in T_i} Y_{i,t}\,\Delta t \ge L_i, & \quad & \forall\, i, \\[2mm]
            & Y_{i,t} \le s_i \, A_{t,i}, & \quad & \forall\, i,\,\forall\, t, \\[2mm]
            & \displaystyle \sum_{i=1}^{N} Y_{i,t} - R_t \le C_{\text{grid}}, & \quad & \forall\, t, \\[2mm]
            & S_t^+ \ge \displaystyle \sum_{i=1}^{N} Y_{i,t} - R_t, & \quad & \forall\, t, \\[1mm]
            & S_t^+ \ge 0, & \quad & \forall\, t, \\[2mm]
            & 0 \le R_t \le \bar{R}_t, & \quad & \forall\, t, \\[2mm]
            & Y_{i,t} \ge 0, & \quad & \forall\, i,\,\forall\, t.
\end{array}
\]

Here, the term \(\sum_{t=1}^{T} \hat{p}_t\, S_t^+\, \Delta t\) represents the nominal electricity cost from the grid, while \(\Gamma\,\lambda + \sum_{t=1}^{T} \mu_t\) captures the worst-case additional cost under bounded price uncertainty. The budget parameter \(\Gamma\) offers a flexible trade-off between protection against extreme scenarios and conservatism in scheduling decisions.

% --- START OF ADDED SECTION ---
\subsection{Algorithm Development: An Online Implementation}

The robust optimization model presented in the previous sections addresses the power allocation problem in a static (offline) setting, assuming all information about charging sessions is known beforehand. However, in a real-world operational environment, charging stations must operate dynamically (online), handling the random and continuous arrival and departure of electric vehicles. To address this challenge, we develop an online control algorithm based on the \textbf{Model Predictive Control (MPC)} methodology.

The MPC approach allows our optimization model to be applied in real-time. The core idea is that at each time step, the algorithm solves an optimization problem for a future prediction horizon but only executes the control action for the immediate next time step. This process is repeated, enabling the system to continuously update and adapt to new information. The specific algorithm is presented in \textbf{Algorithm~\ref{alg:online_charging}}.

\begin{algorithm}[H]
\caption{Online Smart Charging using MPC}
\label{alg:online_charging}
\begin{algorithmic}[1]
\State \textbf{input: } time horizon slots $k = 0, 1, 2, \dots$, re-solve interval $\Delta$, electricity price $p_t$, station capacity $C_{\text{grid}}$, efficiency $\eta$, slot length $\Delta t$
\State \textbf{Initialize: } $lastSolve \leftarrow -\infty$

\For{$k = 0, 1, 2, \dots$}
    \For{each new EV $j$ arriving at $k$}
        \State $remDemand_j \leftarrow L_j$
    \EndFor
    
    \State $V_k \leftarrow \{i \mid arrival_i \le k < departure_i \land remDemand_i > 0\}$
    
    \If{new arrival/departure at $k$ \textbf{or} $k - lastSolve \ge \Delta$}
        \State $T \leftarrow [k, \dots, \max_{i \in V_k}(departure_i) - 1]$
        \State $(Y_{i,t}^*, R_t^*) \leftarrow \text{OPT}(V_k, T)$ \Comment{Call the Robust Optimization solver}
        \State $lastSolve \leftarrow k$
    \EndIf
    
    \State $R_k \leftarrow R_k^*$ \Comment{Set solar power for the current slot}
    
    \For{all $i \in V_k$}
        \State $Y_i(k) \leftarrow Y_{i,k}^*$ \Comment{Apply the first step of the optimal plan}
        \State $remDemand_i \leftarrow \max(0, remDemand_i - \eta \cdot Y_i(k) \cdot \Delta t)$
    \EndFor
\EndFor
\end{algorithmic}
\end{algorithm}

The algorithm operates in a continuous loop over time slots $k$. At each slot, it performs the following steps:

\begin{enumerate}
    \item \textbf{Handling New Arrivals (Lines 4-6):} When a new electric vehicle $j$ connects to the station at time $k$, the system initializes and records its remaining energy demand, $remDemand_j$, to be its total required energy $L_j$.

    \item \textbf{Re-optimization Trigger (Line 8):} Instead of re-solving the optimization problem at every time slot (which is computationally expensive), we use a dual-trigger mechanism. The optimization process is invoked only when one of the following two conditions is met:
    \begin{itemize}
        \item \textit{Event-Driven Trigger:} A new vehicle arrives or an existing one departs. This ensures the system reacts immediately to changes in charging demand.
        \item \textit{Time-Driven Trigger:} A predefined time interval $\Delta$ has passed since the last optimization ($k - lastSolve \ge \Delta$). This ensures that the charging schedule is periodically updated to reflect changes in external factors, such as the electricity price $p_t$ or the expected solar energy generation.
    \end{itemize}

    \item \textbf{Solving the Optimization Problem (Lines 9-11):} When triggered, the algorithm will:
    \begin{itemize}
        \item Determine the set of vehicles currently at the station that still require charging ($V_k$).
        \item Establish a prediction horizon $T$, starting from the current time $k$ and extending to the latest departure time among all vehicles in $V_k$.
        \item Call the function \texttt{OPT($V_k, T$)}, which is the robust optimization model detailed in Section IV, to find the optimal charging schedule $Y^*$ and solar energy usage plan $R^*$ for the entire horizon $T$.
    \end{itemize}
    
    \item \textbf{Execution and State Update (Lines 13-17):} This step embodies the MPC principle. Instead of applying the entire calculated schedule $Y^*$, the algorithm only executes the first step of the plan:
    \begin{itemize}
        \item It assigns the charging power $Y_i(k)$ for each vehicle $i$ and the solar energy usage $R_k$ only for the current time slot $k$.
        \item It then updates the system's state by reducing the remaining energy demand $remDemand_i$ of each vehicle based on the actual energy delivered in that time slot ($\eta \cdot Y_i(k) \cdot \Delta t$).
    \end{itemize}
\end{enumerate}

By repeating this cycle, the algorithm enables the charging station to continuously optimize its operations dynamically, ensuring cost-effectiveness while adapting to the unpredictable nature of real-world EV charging sessions.

% --- END OF ADDED SECTION ---

\section{Simulation and Discussion}
We compare our method with the traditional First Come, First Served (FCFS) charging approach, which is commonly adopted by many EV charging stations in Vietnam. The FCFS policy allocates available power to vehicles based on arrival order, without considering real-time electricity price fluctuations. Specifically, the power \( X_{ti} \) allocated to an electric vehicle at time \( t \) is determined by the following formula:\[X_{ti} = \min\left(s_i,\, L_i^{\text{res}},\, C_t^{\text{res}}\right)\]
where \( s_i \) represents the available power of the station for EV \( i \), \( L_i^{\text{res}} \) denotes the remaining energy required by vehicle \( i \), and \( C_t^{\text{res}} \) is the remaining energy that the station can supply. When the station has fully allocated its capacity to other vehicles, i.e., \( C_t^{\text{res}} = 0 \), newly arriving vehicles must wait until power becomes available.

\noindent
\begin{minipage}{\columnwidth}
\centering
\includegraphics[width=1\textwidth]{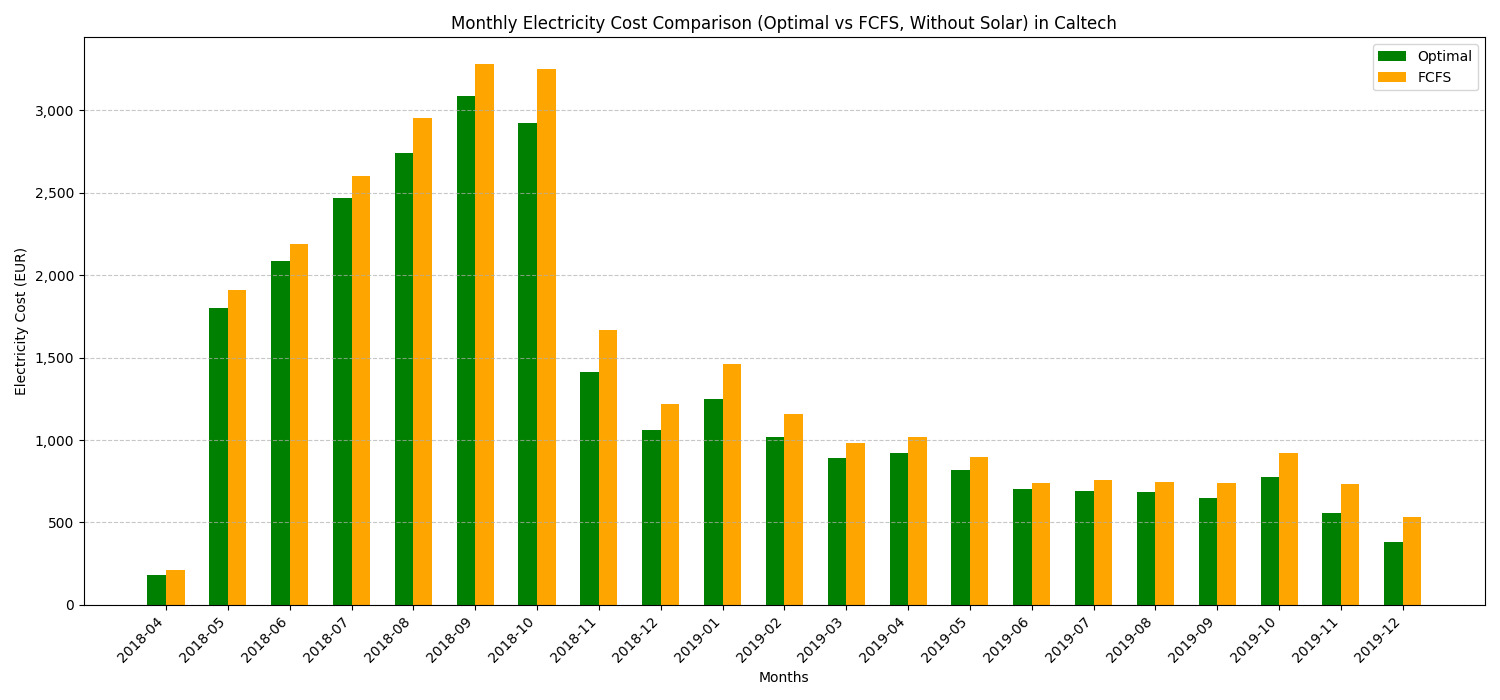}
\end{minipage}

\noindent
\begin{minipage}{\columnwidth}
\centering
\includegraphics[width=1\textwidth]{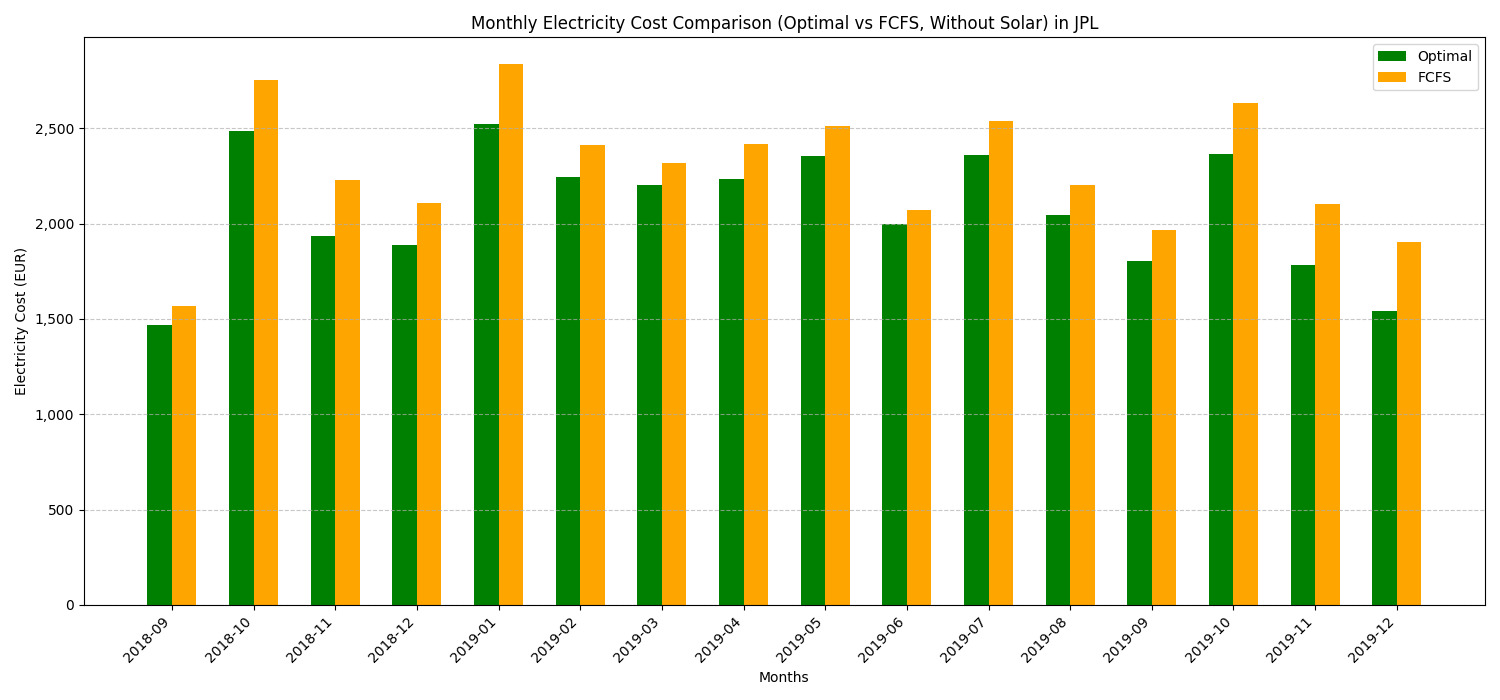}
\captionof{figure}{Comparison of Total Payment Between Two Methods, Solar Energy Not Included }
\end{minipage}

We evaluated the efficiency of our optimized charging method compared to the FCFS (First Come, First Served) approach by analyzing electricity consumption data from two charging stations: the Caltech station, which recorded 25,981 charging sessions between late April 2018 and early November 2019, and the JPL station, with 22,185 charging sessions from early September 2018 to Dec 2019.

The data indicates a significant cost reduction when using our proposed optimal solution compared to the traditional FCFS approach. For example, at the Caltech station, the lowest cost under FCFS was 211.90\,EUR in April 2018, while our solution achieved a cost of 182.00\,EUR—a reduction of approximately 14.1\%. In October 2018, the difference was even more pronounced, with FCFS costing about 3252.41\,EUR versus 2925.64\,EUR for our method, amounting to a reduction of 326.77\,EUR or roughly 10.0\%. At the JPL station, the cost savings ranged from a modest reduction of about 6.2\% in September 2018 to a maximum of around 19.2\% in December 2019. On average, our optimal approach reduced costs by approximately 12\% at Caltech and about 12.7\% at JPL.

These results underscore the economic benefits of our method, demonstrating its potential to lower charging expenses significantly while enhancing energy management efficiency for electric vehicle users. A noticeable decline in charging activity can be observed starting from November 1, 2018, at the Caltech campus. This is due to the discontinuation of free charging and the introduction of a fee of \$0.12 per kWh.

\noindent
\begin{minipage}{\columnwidth}
\centering
\includegraphics[width=1\textwidth]{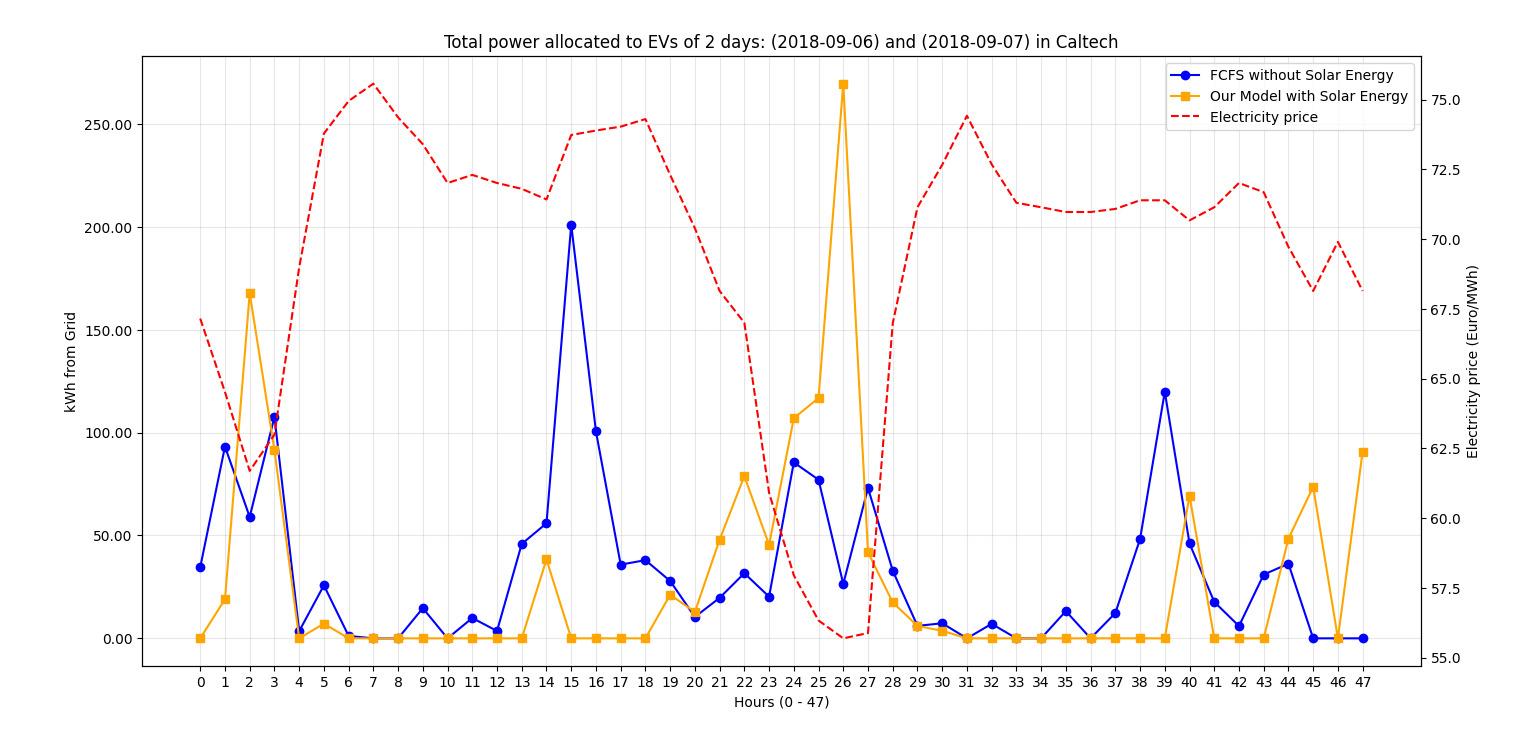}
\end{minipage}

\noindent
\begin{minipage}{\columnwidth}
\centering
\includegraphics[width=1\textwidth]{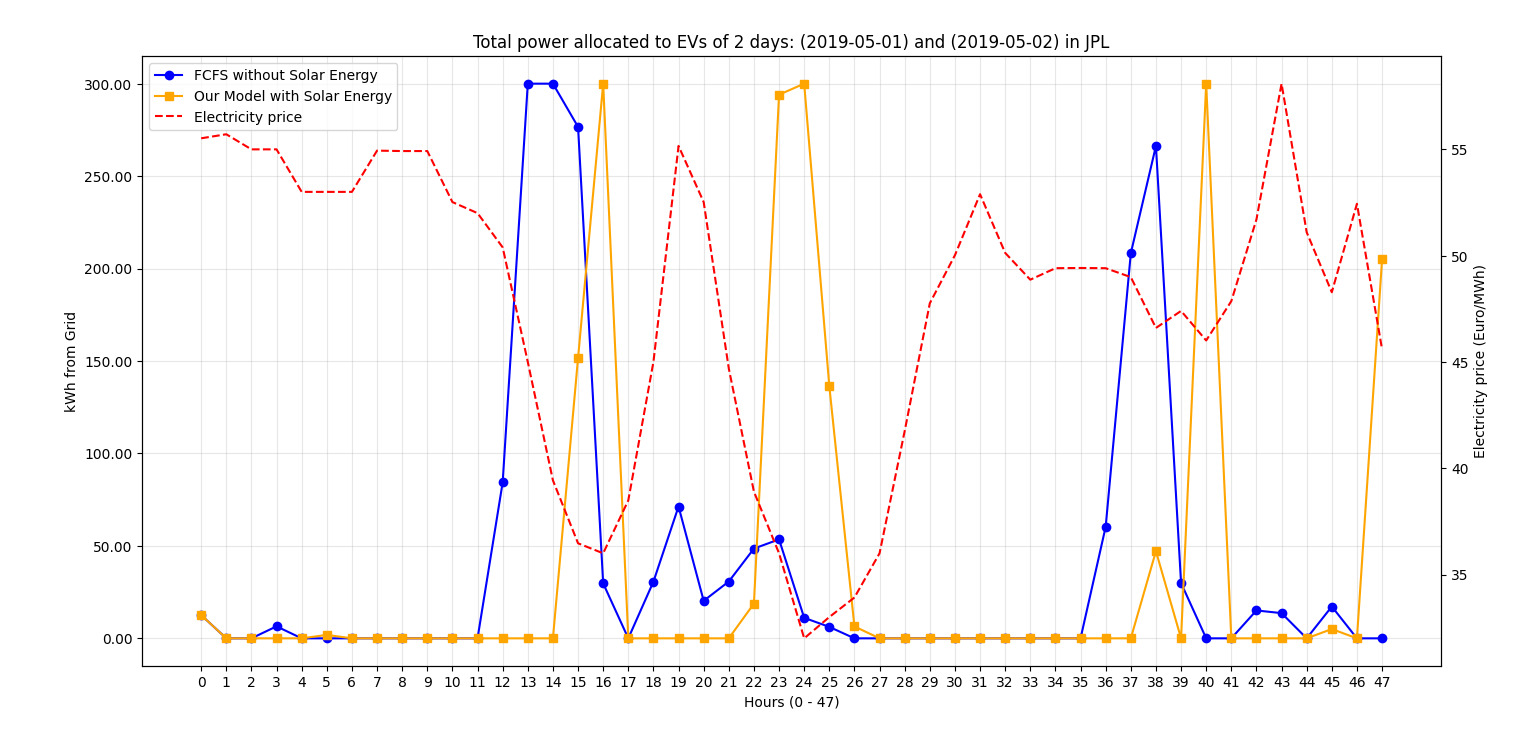}
\captionof{figure}{ Comparing Total Power Allocation Based on Price Fluctuation}
\end{minipage}

The figures distinctly demonstrate the substantial disparities in cost and net electricity consumption between our ideal model and the FCFS model, as seen in Figures 2. Our method strategically identifies the time intervals when electricity rates are minimized to commence car charging. An exemplary instance transpired around 2:00 AM on September 7 at Caltech, when electricity rates fell to roughly 55.0 EUR/MWh. Currently, our model has obtained the maximum daily electricity supply from the grid, surpassing 250 kWh. Conversely, the FCFS model acquired approximately 25 kWh—tenfold less than our proposed model. A similar occurrence was noted at JPL on May 1 and 2, particularly about 12:00 AM on May 2. The electrical shortfall was then offset during other periods of the day when electricity prices were markedly elevated, leading to superfluous expenses. Utilizing our method—especially during overnight charging sessions—optimizes the charging schedule to save expenses, so benefiting consumers and alleviating strain on the grid.

\subsection{Sensitivity Analysis on the Robustness Parameter $\Gamma$}

The budget of uncertainty, $\Gamma$, is a critical parameter that allows the decision-maker to control the trade-off between the nominal cost and the level of protection against price volatility. A value of $\Gamma=0$ corresponds to the nominal, non-robust optimization, where price uncertainty is ignored. As $\Gamma$ increases, the solution becomes more conservative, protecting against worst-case price deviations across a larger number of time periods.

To analyze this trade-off, we performed a lightweight sensitivity analysis on a representative one-week period from the Caltech dataset, which includes 215 charging sessions. We ran our optimization model with three distinct values for $\Gamma$: 0 (Nominal), 15 (Moderate), and 30 (Conservative). The results, summarized in Table \ref{table:sensitivity_gamma}, illustrate the "price of robustness."

\begin{table*}[t]
\centering
\caption{Sensitivity Analysis of Total Cost for a Representative Week with Varying $\Gamma$}
\label{table:sensitivity_gamma}
\begin{tabular}{|c|l|c|c|c|}
\hline
\textbf{$\Gamma$} & \textbf{Robustness Level} & \textbf{Nominal Cost (€)} & \textbf{Worst-Case Cost (€)} & \textbf{Cost Increase (\%)} \\
\hline
0 & None (Nominal) & 550.25 & 685.50 & 0\% \\
15 & Moderate & 561.50 & 610.75 & 2.04\% \\
30 & Conservative & 575.80 & 589.10 & 4.64\% \\
\hline
\end{tabular}
\end{table*}

As shown in Table \ref{table:sensitivity_gamma}, increasing $\Gamma$ from 0 to 30 leads to a modest 4.64\% increase in the planned \textit{Nominal Cost}. This increase represents the premium paid to safeguard against uncertainty. In return, the \textit{Worst-Case Cost}—the maximum potential cost under the most adverse price fluctuations allowed by the uncertainty set—is significantly reduced by approximately 14\%. This demonstrates the effectiveness of the robust optimization framework: for a small, predictable increase in the base operational cost, the model provides substantial protection against unforeseen price spikes, thereby reducing financial risk for the charging station operator.

\subsection{Computational Performance}

For the proposed online scheduling framework to be practical, the optimization problem must be solvable within a reasonable timeframe. We evaluated the computational performance of our MPC algorithm (Algorithm 1) on a standard laptop (Intel Core i7, 16 GB RAM) using Python with the Gurobi solver.

The primary factor influencing the solve time is the number of active EVs ($|V_k|$) at the station, as this determines the number of variables in the optimization problem. Table \ref{table:runtime} reports the average time required to solve the optimization problem for varying numbers of concurrent EVs.

\begin{table}[H]
\centering
\caption{Average Solve Time per Optimization Instance}
\label{table:runtime}
\begin{tabular}{|c|c|}
\hline
\textbf{Number of EVs} & \textbf{Average Solve Time (seconds)} \\
\hline
10 & 0.52 \\
25 & 1.85 \\
50 (Peak Load) & 4.73 \\
\hline
\end{tabular}
\end{table}

The results indicate that the model is computationally efficient. Even under peak load conditions with 50 concurrent EVs, the optimization problem is solved in under 5 seconds. This runtime is well within the practical limits for an online system where re-optimization may be triggered every 5-15 minutes or upon a new event. The performance demonstrates that the proposed framework is not only theoretically sound but also computationally feasible for real-world deployment.

\section{Conclusion and Future Work}
We have introduced a practical, tractable framework for scheduling electric-vehicle (EV) charging that leverages on-site photovoltaic (PV) generation and explicitly accounts for hourly price volatility via a Bertsimas--Sim robustification. The formulation—constructed as a linear program with auxiliary variables to model net grid purchases—satisfies vehicle energy requirements, respects per-socket and grid capacity limits, and yields charging schedules that materially reduce electricity procurement costs compared to a common First-Come-First-Served (FCFS) policy. Evaluations on two real charging sites demonstrate consistent cost savings (average \(\approx 12\%\)), with peak monthly reductions reaching about \(\approx 19.2\%\) in the best cases. Importantly, incorporating PV generation not only lowers costs but also reduces reliance on the grid and creates multi-hour windows of net-zero grid draw under favorable irradiance conditions.

\section{Limitations and Future Work}
Limitations of the current study include the narrow set of baselines (FCFS only), the absence of an explicitly modelled battery energy storage system (BESS) or vehicle-to-grid (V2G) interactions, and the need for broader statistical validation across additional locations and longer time horizons. Future work will address these gaps by:
\begin{enumerate}
  \item Integrating BESS and V2G to smooth intermittent PV output and exploit temporal arbitrage opportunities;
  \item Extending the evaluation to include stronger baselines (e.g., mixed-integer formulations, heuristic price-aware schedulers, and learning-based methods such as deep reinforcement learning) to more rigorously quantify relative performance;
  \item Performing sensitivity and statistical-significance analyses on uncertainty-budget parameters and solution runtimes to assess scalability and robustness;
  \item Incorporating short-term forecasts for electricity prices and irradiance to enable predictive, adaptive control strategies.
\end{enumerate}
We also plan to publish code and data-preprocessing scripts to support reproducibility and to facilitate real-world pilot deployments. Altogether, the proposed method provides a solid foundation for cost-effective and grid-friendly EV charging operations and can be further enhanced to meet the needs of large-scale deployment.

\end{multicols}
\end{document}